\pdfoutput=1
\documentclass[a4paper,UKenglish]{lipics-v2019}

\title{Modular Descriptions of Regular Functions}
\author{Paul Gastin}{LSV, ENS Paris-Saclay, CNRS, Universit{\'e}
Paris-Saclay, France}{paul.gastin@ens-paris-saclay.fr}{}{}

\authorrunning{P.\ Gastin} 

\Copyright{Paul Gastin}



\nolinenumbers 
\hideLIPIcs  

\usepackage[pdflatex,recompilepics=false]{gastex}
\gasset{frame=false,Nw=5,Nh=5,loopdiam=4}

\usepackage{amsmath}
\usepackage{amsfonts}
\usepackage{amssymb}

\newcommand{\leftend}{{\vdash}}
\newcommand{\rightend}{{\dashv}}
\newcommand{\A}{\mathcal{A}}
\newcommand{\dom}{\mathsf{dom}}

\begin{document}

\maketitle

\begin{abstract}
  We discuss various formalisms to describe string-to-string transformations.
  Many are based on automata and can be seen as operational descriptions,
  allowing direct implementations when the input scanner is deterministic.
  Alternatively, one may use more human friendly descriptions based on some
  simple basic transformations (e.g., copy, duplicate, erase, reverse) and
  various combinators such as function composition or extensions of regular
  operations.
\end{abstract}

\begin{gpicture}[name=comments,ignore]
  \node[Nmarks=i,iangle=-90](1)(0,0){1}
  \node(2)(-20,0){2}
  \node(3)(20,0){3}
  \drawloop(1){$b \mid b$}
  \drawedge[curvedepth=1](1,2){$\backslash \mid \backslash$}
  \drawedge[curvedepth=1](2,1){$a \mid a$}
  \drawedge[curvedepth=1](1,3){$\% \mid \varepsilon$}
  \drawedge[curvedepth=1](3,1){$\backslash n \mid \backslash n$}
  \drawloop(3){$c \mid \varepsilon$}
\end{gpicture}
\begin{gpicture}[name=increment-left,ignore]
  \node(1)(0,0){1}
  \node(2)(20,0){2}
  \drawloop(1){$1 \mid 0$}
  \drawedge[curvedepth=0](1,2){$0 \mid 1$}
  \drawloop(2){$\begin{array}{c} 0 \mid 0 \\ 1 \mid 1 \end{array}$}
  \node[Nframe=n](11)(-12,0){} \drawedge(11,1){$\mid \varepsilon$}
  \node[Nframe=n](11)(0,-10){} \drawedge(1,11){${}\mid 1$}
  \node[Nframe=n](12)(20,-10){} \drawedge(2,12){${}\mid \varepsilon$}
\end{gpicture}
\begin{gpicture}[name=increment-right,ignore]
  \node(1)(0,0){1}
  \node(2)(20,0){2}
  \drawloop(2){$1 \mid 0$}
  \drawedge[curvedepth=0](1,2){$0 \mid 1$}
  \drawloop(1){$\begin{array}{c} 0 \mid 0 \\ 1 \mid 1 \end{array}$}
  \node[Nframe=n](11)(0,-10){} \drawedge[ELside=r](11,1){${}\mid \varepsilon$}
  \node[Nframe=n](12)(20,-10){} \drawedge[ELside=r](12,2){${}\mid 1$}
  \node[Nframe=n](12)(31,0){} \drawedge[ELside=l](2,12){${}\mid \varepsilon$}
\end{gpicture}
\begin{gpicture}[name=increment-2way,ignore]
  \node[Nmarks=i](0)(0,0){0}
  \node(1)(20,0){1}
  \node(2)(45,0){2}
  \node(3)(70,0){3}
  \node(4)(20,-14){4}
  \node(5)(45,-14){5}
  \node[Nmarks=f](6)(70,-14){6}
  \drawedge[curvedepth=0](0,1){\small$\leftend \mid \leftend,\rightarrow$}
  \drawloop(1){\small$1 \mid \varepsilon,\rightarrow$}
  \drawedge[curvedepth=0](1,2){\small$0 \mid \varepsilon,\leftarrow$}
  \drawloop(2){\small$1 \mid \varepsilon,\leftarrow$}
  \drawedge[curvedepth=0](2,3){\small$\begin{array}{c} 0 \mid 0,\rightarrow \\ 
  \leftend \mid \varepsilon,\rightarrow \end{array}$}
  \drawloop(3){\small$1 \mid 1,\rightarrow$}
  \drawedge[curvedepth=4](3,1){\small$0 \mid \varepsilon,\rightarrow$}
  \gasset{loopangle=-90}
  \drawedge[curvedepth=0,ELside=r](1,4){\small$\rightend \mid \varepsilon,\leftarrow$}
  \drawloop(4){\small$1 \mid \varepsilon,\leftarrow$}
  \drawedge[curvedepth=0,ELside=r](4,5){\small$\begin{array}{c} 0 \mid 1,\rightarrow \\ 
  \leftend \mid 1,\rightarrow \end{array}$}
  \drawloop(5){\small$1 \mid 0,\rightarrow$}
  \drawedge[curvedepth=0,ELside=r](5,6){\small$\rightend \mid \rightend,\rightarrow$}
\end{gpicture}
\begin{gpicture}[name=increment-register-copy,ignore]
  \gasset{ELdist=1.5}
  \node(1)(0,0){1}
  \node[Nframe=n](2)(0,11){}
  \node[Nframe=n](3)(0,-11){}
  \drawedge[ELpos=35](2,1){\small$X:=\varepsilon ; Y:=1$}
  \drawedge[ELpos=65](1,3){\small$Y$}
  \drawloop[loopangle=180](1){\small$1 \mid X:=X1 ; Y:=Y0$}
  \drawloop[loopangle=0](1){\small$0 \mid Y:=X1 ; X:=X0$}
\end{gpicture}
\begin{gpicture}[name=increment-register,ignore]
  \gasset{ELdist=1.5}
  \node[Nframe=n](0)(0,11){}
  \node[Nmarks=f](1)(0,0){1}
  \node(2)(0,-15){2}
  \node[Nframe=n](3)(0,-26){}
  \drawedge[ELpos=35](0,1){\small$X:=\varepsilon ; Y:=\varepsilon ; Z:=\varepsilon$}
  \drawloop[loopangle=180](1){\small$1 \mid X:=X1 ; Y:=Y0 ; Z:=Z$}
  \node[Nframe=n](11)(12,0){\small$Z 1 Y$}

  \drawedge[ELpos=50](1,2){\small$0 \mid Z:=X ; X:=\varepsilon ; Y:=\varepsilon$}
  \drawloop[loopangle=180](2){\small$1 \mid X:=X1 ; Y:=Y0 ; Z:=Z$}
  \drawloop[loopangle=0](2){\small$0 \mid Z:=Z 0 X ; X:=\varepsilon ; Y:=\varepsilon$}
  \drawedge[ELpos=65](2,3){\small$Z 1 Y$}
\end{gpicture}

We investigate string-to-string functions which are ubiquitous.  A
preprocessing that erases comments from a program, or a micro-computation that
replaces a binary string with its increment, or a syntactic fix that reorders
the arguments of a function to comply with a different syntax, are all examples
of string-to-string transformations/functions.  We discuss 
various ways of describing such functions and survey some of the main 
results.

\begin{figure}
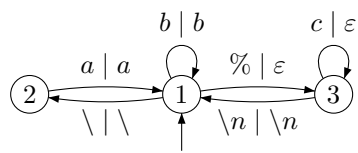

  \centerline{\gusepicture{comments}}
  \caption{A sequential transducer stripping comments from a latex file, where
  $a,b,c\in\Sigma$ are letters from the input alphabet with
  $b\notin\{\backslash,\%\}$ and $c\neq\backslash n$.}
  \protect\label{fig:comments}
\end{figure}

Operationally, we need to parse the input string and to produce an output word.
The simplest such mechanism is to use a deterministic finite-state automaton
(1DFA) to parse the input from left to right and to produce the output along the
way.  These are called \emph{sequential} transducers, or one-way
input-deterministic transducers (1DFT), see e.g.~\cite[Chapter
IV]{Berstel_1979}, \cite[Chapter V]{Sakarovitch09} or \cite{FiliotReynier16}.
Transitions are labelled with pairs $a\mid u$ where $a$ is a letter read from
the input string and $u$ is the word, possibly empty, to be appended to the
output string.  Sequential transducers allow for instance to strip comments from
a latex file, see Figure~\ref{fig:comments}.  Transformations that can be
realized by a sequential transducer are called sequential functions.  A very
important property of sequential functions is that they are closed under
composition.  This can be easily seen by taking a cartesian product of the two
sequential transducers, synchronizing the output of the first transducer with
the input of the second one.  Also, each sequential function $f$ can be realized
with a canonical minimal sequential transducer $\mathcal{A}_f$ which can be
computed from any sequential transducer $\mathcal{B}$ realizing $f$.  As a
consequence, equivalence is decidable for sequential transducers.

\begin{figure}[tbp]
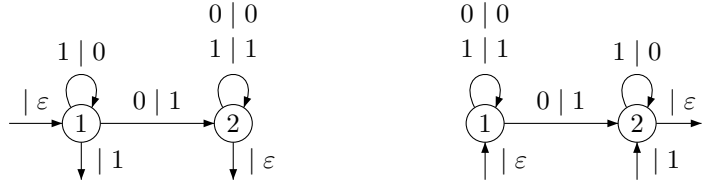

  \centerline{\gusepicture{increment-left}\hfil\gusepicture{increment-right}}
  \caption{Transducers incrementing a binary number.}
  \protect\label{fig:increment}
\end{figure}

With a sequential transducer, it is also possible to increment an integer
written in binary if the string starts with the least significant bit (lsb), see
Figure~\ref{fig:increment} left.  On the other hand, increment is not a
sequential function when the lsb is on the right.  There are two possibilities
to overcome this problem.  

The first solution is to give up determinism when reading the
input string.  One-way input-nondeterministic finite-state transducers (1NFT)
do not necessarily define functions.  It is decidable in PTIME whether a 1NFT
defines a function \cite{Schutzenberger_1975,Gurari_1983}. 
We give a proof below\footnote{
Let $\A$ be a 1NFT with $m$ states.
We show that, if $\A$ is functional on all words of length $\leq 2m^{2}$,
then $\A$ is functional.
\\
Let $w=a_1a_2\ldots a_n\in\dom(\A)$ with $n>2m^{2}$.  By induction, we assume
that $\A$ is functional on all words of length $<n$.  Consider two accepting
runs for $w$: 
$p_0\xrightarrow{a_1}p_1\xrightarrow{a_2}p_2\cdots p_{n-1}\xrightarrow{a_n}p_n$ and
$q_0\xrightarrow{a_1}q_1\xrightarrow{a_2}q_2\cdots q_{n-1}\xrightarrow{a_n}q_n$.
\\
Since $n$ is large enough, we find  
$0\leq i<j<k\leq n$ with $(p_i,q_i)=(p_j,q_j)=(p_k,q_k)$.
We split the input word $w=w_1w_2w_3w_4$ in four factors 
$w_1=a_1\cdots a_i$, $w_2=a_{i+1}\cdots a_j$, $w_3=a_{j+1}\cdots 
a_k$ and $w_4=a_{k+1}\cdots a_n$ and we consider the ouputs $x_1x_2x_3x_4$ and 
$y_1y_2y_3y_4$ of the two accepting runs:
$p_0\xrightarrow{w_1 \mid x_1} 
p_i\xrightarrow{w_2 \mid x_2} 
p_j\xrightarrow{w_3 \mid x_3} 
p_k\xrightarrow{w_4 \mid x_4} p_n$
and
$q_0\xrightarrow{w_1 \mid y_1} 
q_i\xrightarrow{w_2 \mid y_2} 
q_j\xrightarrow{w_3 \mid y_3} 
q_k\xrightarrow{w_4 \mid y_4} q_n$.
\\
The three repeated pairs allow us to consider shortcuts in the accepting paths.
First we skip $w_2w_3$ and we get two accepting runs for the shorter word
$w_1w_4$: $p_0\xrightarrow{w_1 \mid x_1} p_i=p_k\xrightarrow{w_4 \mid x_4} p_n$
and $q_0\xrightarrow{w_1 \mid y_1} q_i=q_k\xrightarrow{w_4 \mid y_4} q_n$.  By
induction, the outputs must be equal: $x_1x_4 = y_1y_4$.  Wlog we assume that
$y_1$ is a prefix of $x_1$ and we obtain $x_1=y_1z$ and $zx_4=y_4$ for some $z$.
\\
Second, we skip $w_3$ and by induction the ouputs on the shorter word
$w_1w_2w_4$ should be equal: $x_1x_2x_4 = y_1y_2y_4$.  Therefore,
$y_1zx_2x_4=y_1y_2zx_4$ and $zx_2=y_2z$.
Similarly, skipping $w_2$ we get
$x_1x_3x_4 = y_1y_3y_4$ and $zx_3=y_3z$.
Finally,
$x_1x_2x_3x_4 = y_1 z x_2x_3x_4 = y_1 y_2 z x_3x_4
= y_1 y_2 y_3 z x_4 = y_1 y_2 y_3 y_4$.
Hence, $\A$ is functional on $w$
}
which is mostly inspired from \cite[Chapter IV]{Berstel_1979}.

We are interested in \emph{functional} 1NFT (f1NFT).  This is in particular the
case when the transducer is input-\emph{unambiguous}.  Actually, one-way,
input-unambiguous, finite-state transducers (1UFT) have the same expressive
power as f1NFT \cite{Weber_1995}.  We prove this result below when discussing
regular look-ahead.  For instance, increment with lsb on the right is realized
by the 1UFT on Figure~\ref{fig:increment} right.  Transformations
realized by f1NFT are called rational functions.  They are easily closed under
composition.  The equivalence problem is undecidable for 1NFT
\cite{Griffiths_1968} but decidable in PTIME for f1NFT
\cite{Schutzenberger_1975,Gurari_1983}.  This follows directly from the
decidability of the functionality of 1NFTs: consider two f1NFTs $\A_1$ and
$\A_2$, first check whether $\dom(\A_1)=\dom(\A_2)$, then check whether
$\A_1\uplus\A_2$ is functional.  It is also decidable in PTIME whether a f1NFT
defines a sequential function, i.e., whether it can be realized by a 1DFT
\cite{Choffrut_1977,Weber_1995}.

Interestingly, any rational function $h$ can be written as $r\circ g\circ r\circ
f$ where $f,g$ are sequential functions and $r$ is the \emph{reverse} function
mapping $w=a_1a_2\cdots a_n$ to $w^{r}=a_n\cdots a_2a_1$ \cite{Elgot_1965}.  We
provide a sketch of proof below.\footnote{
Assume that $h$ is realized by a 1UFT $\mathcal{B}$.  Consider the unique
accepting run $q_0\xrightarrow{a_1\mid u_1}q_1 \cdots
q_{n-1}\xrightarrow{a_n\mid u_n}q_n$ of $\mathcal{B}$ on some input word
$w=a_1\cdots a_n$.  We have $h(w)=u_1\cdots u_n$.  
\\
Let $\mathcal{A}$ be the DFA obtained with the subset construction applied to
the input NFA induced by $\mathcal{B}$.  Consider the run
$X_0\xrightarrow{a_1}X_1\cdots X_{n-1}\xrightarrow{a_n}X_{n}$ of $\mathcal{A}$
on $w$.  We have $q_i\in X_i$ for all $0\leq i\leq n$.  The first sequential
function $f$ adorns the input word with the run of $\mathcal{A}$:
$f(w)=(X_0,a_1)\cdots(X_{n-1},a_n)$.  
\\
The sequential transducer $\mathcal{C}$ realizing $g$ is defined as follows.
For each state $q$ of $\mathcal{B}$ there is a transition
$\delta=q\xrightarrow{(X,a)}p$ in $\mathcal{C}$ if there is a unique $p\in X$
such that $\delta'=p\xrightarrow{a}q$ is a transition in $\mathcal{B}$.
Moreover, if $\delta'$ outputs $u$ in $\mathcal{B}$ then $\delta$ outputs
$u^{r}$ in $\mathcal{C}$.  
\\
Notice that $q_n\xrightarrow{(X_{n-1},a_n)\mid
u_n^{r}}q_{n-1}\cdots q_1\xrightarrow{(X_0,a_1)\mid u_1^{r}}q_0$ is a run of
$\mathcal{C}$ producing $u_n^{r}\cdots u_1^{r}=h(w)^{r}$.  
\\
The result follows.}

In classical automata, whether or not a transition can be taken only depends on
the input letter being scanned.  This can be enhanced using regular look-ahead
or look-behind.  For instance, the f1NFT on the right of
Figure~\ref{fig:increment} can be made deterministic using regular look-ahead.
In state 1, when reading digit 0, we move to state 2 if the suffix belongs to
$1^{*}$ and we stay in state 1 otherwise, i.e., if the suffix belongs to
$1^{*}0\{0,1\}^{*}$.  Similarly, we choose to start in the initial state $2$
(resp.\ 1) if the word belongs to $1^{*}$ (resp.\ $1^{*}0\{0,1\}^{*}$).  More
generally, any f1NFT can easily be made deterministic using regular look-ahead:
we consider an arbitrary total order $<$ on the set of states of the f1NFT and
we select the \emph{least} accepting path for the lexicographic ordering.  If
from state $p$ reading $a$ we have the choice between several transitions
leading to states $q_1<q_2<q_3\cdots$, we select the least $i$ such that the
suffix can be accepted from $q_i$.  This query is indeed regular.  We deduce
that regular look-ahead increases the expressive power of one-way deterministic
transducers.

Notice that a one-way transducer which is deterministic thanks to regular
look-ahead can be easily transformed into a 1UFT. For instance, if a
non-deterministic choice between $p\xrightarrow{a,L_1}q_1$ and
$p\xrightarrow{a,L_2}q_2$ is resolved by the disjoint regular look-ahead $L_1$
and $L_2$, then the 1UFT goes to $q_1$ (or $q_2)$ and spans a copy of the
automaton for $L_1$ (or $L_2)$ to check that the suffix satisfies the correct
look-ahead.  We have actually proved that f1NFT and 1UFT have the same
expressive power: starting with a f1NFT, we get a deterministic transducer using
regular look-ahead, then we turn it into a 1UFT.

\begin{figure}[tbp]
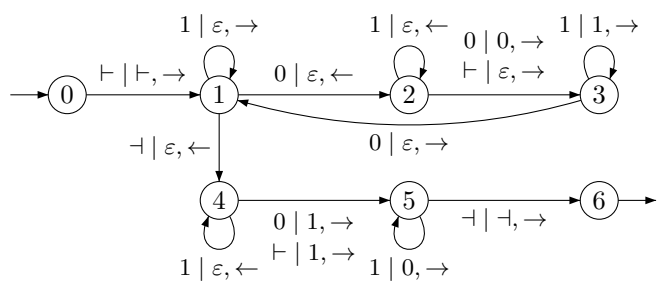

  \centerline{\gusepicture{increment-2way}}
  \caption{Two-way transducer incrementing a binary number.}
  \protect\label{fig:increment-2way}
\end{figure}

\medskip
Remember that increment with lsb on the right is not a sequential function. The 
first solution was to use f1NFT or 1UFT as in Figure~\ref{fig:increment} right.
The other solution is to keep input-determinism but to allow the transducer to
move its input head in both directions, i.e., left or right (two-way).  So we
consider two-way input-deterministic finite-state transducers (2DFT)
\cite{Aho_1970}.  To realize increment of binary numbers with the lsb on the
right with a 2DFT, one has to locate the last 0 digit, replace it with 1, keep
unchanged the digits on its left and replace all 1's on its right with 0's.
This is realized by the 2DFT of Figure~\ref{fig:increment-2way}.  We use
$\leftend,\rightend\notin\Sigma$ for the end-markers so the input tape contains
$\leftend w \rightend$ when given the input word $w\in\Sigma^{*}$.

\medskip
Transformations realized by 2DFTs are called regular functions.  They form a
very robust class.  Remarkably, regular functions are closed under composition
\cite{Chytil1977}, which is now a non trivial result.  Actually, a 2DFT can be
transformed into a \emph{reversible} one of exponential size
\cite{dartois_et_al:Icalp2017}.  In a reversible transducer, computation steps
can be deterministically reversed.  As a consequence, the composition of two
2DFTs can be achieved with a single exponential blow-up.
Also, contrary to the one-way case, input-nondeterminism does not add expressive
power as long as we stay functional: given a f2NFT, one may construct an
equivalent 2DFT \cite{Engelfriet_2001}.  Similarly, regular look-ahead and
look-behind do not increase the expressive power of regular functions
\cite{Engelfriet_2001}.
Moreover, the equivalence problem for regular functions is still
decidable~\cite{Culik_1986}.

Regular functions are also those that can be defined with MSO transductions
\cite{Engelfriet_2001}, but we will not discuss this here.

\begin{figure}[tbp]
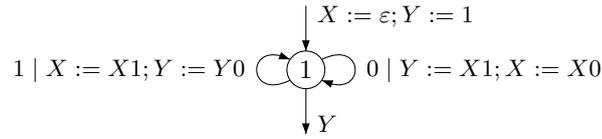

  \centerline{\gusepicture{increment-register-copy}}
  \caption{One-way register transducer incrementing a binary number.}
  \protect\label{fig:increment-register-copy}
\end{figure}

By using registers, we obtain yet another formalism defining string-to-string
transformations.  For instance incrementing a binary number with lsb on the
right is realized by the one-way register transducer on
Figure~\ref{fig:increment-register-copy}.  It uses two registers $X,Y$
initialized with the empty string and 1 respectively and updated while reading
the binary number.  Register $X$ keeps a copy of the binary number read so far,
while $Y$ contains its increment.  The final output of the transducer is the
string contained in register $Y$.  This register automaton is a special case of
``\emph{simple programs}'' defined in \cite{Chytil1977}.  In these simple
programs, a register may be reset to the empty string, copied to another
register, or updated by appending a finite string.  The input head is two-way
and most importantly simple programs may be composed.  Simple programs coincide
in expressive power with 2DFTs \cite{Chytil1977}, hence define once again the
class of regular functions.

\begin{figure}[tbp]
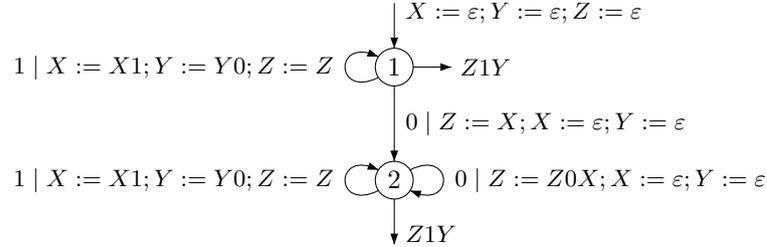

  \centerline{\gusepicture{increment-register}}
  \caption{Streaming string transducer incrementing a binary number.}
  \protect\label{fig:increment-register}
\end{figure}

Notice that when reading digit 0, the transducer of
Figure~\ref{fig:increment-register-copy} copies the string stored in $X$ into
$Y$ without resetting $X$ to $\varepsilon$.  By restricting to one-way register
automata with copyless updates (e.g., not of the form $Y:=X1 ; X:=X0$ where the
string contained in $X$ is duplicated) but allowing concatenation of registers
in updates (e.g., $Z := Z0X ; X:=\varepsilon$), we obtain another kind of
machines, called copyless streaming string transducers (SST), once again
defining the same class of regular functions~\cite{Alur-Icalp11}.  Continuing
our example, incrementing a binary number with lsb on the right can be realized
with the SST on Figure~\ref{fig:increment-register}.  It uses three registers
$X,Y,Z$ initialized with the empty string and updated while reading the binary
number.  Register $X$ keeps a copy of the last sequence of 1's while register
$Y$ contains a sequence of 0's of same length.  Now register $Z$ keeps a copy of
the input read so far up to, and excluding, the last 0.  Hence, the increment of
the binary number read so far is given by $Z1Y$ which is the final output of the
transducer.
If the input number is $1^{n}$ then the computation ends in state 1 with
$Y=0^{n}$ and $Z=\varepsilon$.  Hence the final ouput is $Z1Y=10^{n}$.
Similarly, if the input number is of the form $w01^{n}$ then the run ends in
state $2$ with $Z=w$ and $Y=0^{n}$: the final output is $Z1Y=w10^{n}$.

The above machines provide a way of describing string-to-string transformations
which is not modular.  Describing regular functions in such devices is
difficult, and it is even more difficult to understand what is the function
realized by a 2DFT or an SST. We discuss now more compositional and modular
descriptions of regular functions. Such a formalism, called regular list 
functions, was described in \cite{BojanczykDaviaudKrishna_2018}. It is based on 
function composition together with some natural functions over lists such as 
reverse, append, co-append, map, etc. Here we choose to look at combinators 
derived from regular expressions.

The idea is to start from basic functions, e.g.,
$(1\mid 0)$ means ``read 1 and output 0'', and to apply simple combinators
generalizing regular expressions
\cite{AlurFreilichRaghothaman14,AlurDAntoniRaghothaman_2015,DGK-lics18,BR-DLT18}.
For instance, using the Kleene iteration, $(1\mid 0)^{*}$ describes a function
which replaces a sequence of 1's with a sequence of 0's of same length.
Similarly, $\mathsf{copy}:= ((0\mid 0)+(1\mid 1))^{*}$ describes a regular
function which simply copies an input binary string to the output.  Now,
incrementing a binary number with lsb on the right is described with the
expression $\mathsf{increment0}:=\mathsf{copy}\cdot(0\mid 1)\cdot(1\mid 0)^{*}$,
assuming that the input string contains at least one 0 digit.  If the input
string belongs to $1^{*}$, we may use the expression
$\mathsf{increment1}:=(\varepsilon\mid 1)\cdot(1\mid 0)^{*}$.  Notice that such
a regular transducer expression (RTE) defines simultaneously the \emph{domain}
of the regular function as a regular expression, e.g.,
$\dom(\mathsf{increment0})=(0+1)^{*}01^{*}$, and the output to be
produced.  The input regular expression explains how the input should be parsed.
If the input regular expression is ambiguous, parsing the input word is not
unique and the expression may be non functional.  For instance,
$\mathsf{copy}\cdot(1\mid 0)^{*}$ is ambiguous.  The input word $w=1011$ may
be parsed as $10\cdot 11$ or $101\cdot 1$ or $1011\cdot\varepsilon$ resulting in
the outputs $1000$ or $1010$ or $1011$ respectively. On the other end, 
$\mathsf{increment}:=\mathsf{increment0}+\mathsf{increment1}$ has an 
unambiguous input regular expression.

\emph{Simple} RTEs are defined by the syntax
$$
f,g ::= (u,v) \mid f+g \mid f\cdot g \mid f^{*}
$$
where $u$ is a finite input word, $v$ is a finite output word, and the rational
operations should be unambiguous.  For instance, $f^{*}$ is unambiguous if for
all input words $w$, there is at most one factorization $w=u_1u_2\cdots u_n$
with $u_i\in\dom(f)$.  Simple RTEs define precisely the rational functions
(f1NFT or 1UFT).  This follows from a more general result: the equivalence of
weighted automata and rational series, usually referred to as the 
Kleene-Schützenberger theorem \cite{Schutzenberger61-ic}, applied to the 
semiring of rational languages and restricted to unambiguous weighted automata.

A 2DFT may easily duplicate the input word, defining the function $w\mapsto
w\#w$, which cannot be computed with a sequential transducer or a f1NFT. In
addition to the classical regular combinators ($+$ for disjoint union, $\cdot$
for unambiguous concatenation or Cauchy product, ${}^{*}$ for unambiguous Kleene
iteration), we add the Hadamard product $(f\odot g)(w)=f(w)\cdot g(w)$ where the
input word is read twice, first producing the output computed by $f$ then the
output computed by $g$.  Hence the function duplicating its input can be simply
written as
$\mathsf{duplicate}:=(\mathsf{copy}\cdot(\varepsilon\mid\#))\odot\mathsf{copy}$.
The Hadamard product also allows to exchange two strings $u\#v \mapsto vu$
where $u,v\in\{0,1\}^{*}$.  Let $\mathsf{erase}:=((0\mid \varepsilon)+(1\mid
\varepsilon))^{*}$ and
$$
\mathsf{exchange}:=
\Big( \mathsf{erase}\cdot(\#\mid\varepsilon)\cdot\mathsf{copy}) \Big)
\odot
\Big( \mathsf{copy}\cdot(\#\mid\varepsilon)\cdot\mathsf{erase} \Big) \,.
$$

A 2DFT may also scan its input back and forth in pieces. This was used in 
the 2DFT of Figure~\ref{fig:increment-2way} to locate the last 0 of the input. 
This is also needed to realize the regular function $h$ defined by
$$
h\colon u_1\#u_2\#u_3\#\cdots u_n\# \mapsto u_2u_1\#u_3u_2\#\cdots u_nu_{n-1}\#
$$
where $u_1,\ldots,u_n\in\{0,1\}^{*}$ and $n>1$.  It is easy to build a 2DFT
realizing $h$, but this regular function cannot be expressed using the regular
combinators $+$, $\cdot$, $*$, $\odot$.  On the other hand, we show that $h$ can
be expressed with the help of composition.  First, we iterate the function
$\mathsf{duplicate}$ on a $\#$-separated sequence of binary words with the RTE
$f:=(\mathsf{duplicate}\cdot(\#\mid\#))^{*}$.  We have
$$
f\colon u_1\#u_2\#u_3\#\cdots u_n\# \mapsto 
u_1\#u_1\#u_2\#u_2\#u_3\#u_3\#\cdots u_n\#u_n\#
$$
when $u_1,\ldots,u_n$ are binary strings.
Next, we erase the first $u_1$ and the last $u_n$ and we exchange the remaining 
consecutive pairs with the RTE
$$
g:=\mathsf{erase}\cdot(\#\mid\varepsilon)\cdot(\mathsf{exchange}\cdot(\#\mid\#))^{*}
\cdot\mathsf{erase}\cdot(\#\mid\varepsilon)\,.
$$
We have $g\circ f\colon u_1\#u_2\#u_3\#\cdots u_n\# \mapsto
u_2u_1\#u_3u_2\#\cdots u_nu_{n-1}\#$.  Hence, $h=g\circ f$.

Another crucial feature of 2DFTs is their ability to reverse the input, i.e., to 
implement the function $\mathsf{reverse}\colon a_1a_2\cdots a_n \mapsto 
a_n\cdots a_2a_1$. We add the basic function $\mathsf{reverse}$ to our 
expressions and we obtain RTEs with composition, Hadamard product and reverse 
(chr-RTE) following the syntax:
$$
f,g ::= \textcolor{blue}{\mathsf{reverse}} 
\mid (u,v) 
\mid f+g \mid f\cdot g \mid f^{*}
\mid \textcolor{blue}{f\odot g} \mid \textcolor{blue}{f\circ g} 
$$
where $u$ is a finite input word, $v$ is a finite output word, and the rational
operations $+$, $\cdot$, $*$ should be unambiguous.  It turns out that regular
functions (2DFTs) are exactly those that can be described with chr-RTEs.
Further, we may remove the Hadamard product if we provide $\mathsf{duplicate}$
as a basic function.  Indeed, we can easily check that $f\odot g=
(f\cdot(\#\mid\varepsilon)\cdot g)\circ\mathsf{duplicate}$.  We obtain RTEs with
composition, duplicate and reverse (cdr-RTE) following the syntax:
$$
f,g ::= \textcolor{blue}{\mathsf{reverse}} 
\mid \textcolor{blue}{\mathsf{duplicate}} 
\mid (u,v) 
\mid f+g \mid f\cdot g \mid f^{*}
\mid \textcolor{blue}{f\circ g} \,.
$$
Once again, cdr-RTEs define exactly the class of regular functions.  We believe
that both chr-RTE and cdr-RTE form very convenient, compositional and modular
formalisms for defining regular functions.

\medskip
An alternative solution to the fact that the regular function $h$ defined above
cannot be described using the regular combinators $+$, $\cdot$, $*$, $\odot$ was
proposed in \cite{AlurFreilichRaghothaman14}.  Instead of using composition,
they introduced a 2-chained Kleene iteration: $[K,f]^{2+}$ first
\emph{unambiguously} parses an input word as $w=u_1 u_2\cdots u_n$ with
$u_1,\ldots,u_n\in K$ and then apply $f$ to all consecutive pairs of factors,
resulting in the output $f(u_1u_2)f(u_2u_3)\cdots f(u_{n-1}u_n)$.  For instance,
with the functions defined above, we can easily check that $h=[K,f]^{2+}$
with $K=\{0,1\}^{*}\#$ and $f:=\mathsf{exchange}\cdot(\#\mid\#)$.

We show that the 2-chained Kleene iteration $[K,f]^{2+}$ can be expressed if we
allow composition of functions in addition to the regular combinators $+$,
$\cdot$, $*$, $\odot$.  First, consider an \emph{unambiguous} regular expression
for the regular language $K$ in which we replace each atomic letter $a$ with
$(a\mid a)$.  We obtain a simple RTE $f_K$ with domain $K$ and which is the
identity on its domain $K$.  Now consider the function $g_K$ defined by the
simple RTE $g_K=(f_K\cdot(\varepsilon\mid\#))^{*}$.  When an input word $w$ can
be \emph{unambiguously} parsed as $w=u_1 u_2\cdots u_n$ with $u_1,\ldots,u_n\in
K$, we get $g_K(w)=u_1\#u_2\#\cdots u_n\#$.
As above, we consider the function $g:=(\mathsf{duplicate}\cdot(\#\mid\#))^{*}$
so that $(g\circ g_K)(w)=u_1\#u_1\#u_2\#u_2\#u_3\#u_3\#\cdots u_n\#u_n\#$.  With
a further composition, we erase the first $u_1$ and the last $u_n$ and we apply 
$f$ to  the remaining consecutive pairs with the RTE
$$
h:=\mathsf{erase}\cdot(\#\mid\varepsilon)\cdot
(f\circ(\mathsf{copy}\cdot(\#\mid\varepsilon)\cdot\mathsf{copy}\cdot(\#\mid\varepsilon)))^{*}
\cdot\mathsf{erase}\cdot(\#\mid\varepsilon)\,.
$$
We obtain $[K,f]^{2+}=h\circ g\circ g_K$. Therefore, regular functions 
described by RTEs using combinators 
$+$, $\cdot$, $*$, $\odot$, $2+$ can be expressed with ch-RTEs using 
combinators $+$, $\cdot$, $*$, $\odot$, $\circ$ or cd-RTEs using 
$\mathsf{duplicate}$ instead of the Hadamard product.

\medskip
Since the regular function $\mathsf{reverse}$ cannot be expressed with the
regular combinators $+$, $\cdot$, $*$, $\odot$ and $2{+}$, reversed versions of
Kleene star and 2-chained Kleene iteration were also introduced in
\cite{AlurFreilichRaghothaman14}.  The \emph{reversed} Kleene star $r\text{-}*$
parses the input word from left to right but produces the output in reversed
order.  For instance, $f^{r\text{-}*}(w)=f(u_n)\cdots f(u_2)f(u_1)$ if the input
word is unambiguously parsed as $w=u_1 u_2\cdots u_n$ with $u_i\in\dom(f)$.
Hence, reversing a binary string is described with the expression $((0\mid
0)+(1\mid 1))^{r\text{-}*}$.

Conversely, the reversed Kleene star can be expressed with the basic function
$\mathsf{reverse}$ and composition:
$f^{r\text{-}*}=(f\circ\mathsf{reverse})^{*}\circ\mathsf{reverse}$.  Indeed,
assume that an input word is unambiguously parsed as $w=u_1u_2\cdots u_n$ when
applying $f^{r\text{-}*}$ resulting in $f(u_n)\cdots f(u_2)f(u_1)$.  Then,
$\mathsf{reverse}(w)=w^{r}$ is unambiguously parsed as $u_n^{r}\cdots
u_2^{r}u_1^{r}$ when applying $(f\circ\mathsf{reverse})^{*}$.  The result
follows since $(f\circ\mathsf{reverse})(u^{r})=f(u)$.

There is also a reversed version of the two-chained Kleene iteration.  With the
above notation, we get $[K,h]^{r\text{-}2+}(w)=h(u_{n-1}u_n)\cdots
h(u_2u_3)h(u_1u_2)$ when the input word can be \emph{unambiguously} parsed as
$w=u_1 u_2\cdots u_n$ with $u_1,\ldots,u_n\in K$.

Once again, we obtain an equivalent formalism for describing regular functions:
the regular transducer expressions using $+$, $\cdot$, $\odot$, $*$,
$r\text{-}*$, $2+$, $r\text{-}2+$ as combinators
\cite{AlurFreilichRaghothaman14,AlurDAntoniRaghothaman_2015,DGK-lics18,BR-DLT18}:
$$
f,g ::= (u,v) 
\mid f+g \mid f\cdot g \mid f^{*}
\mid \textcolor{blue}{f\odot g}
\mid \textcolor{blue}{f^{r\text{-}*}}
\mid \textcolor{blue}{[K,f]^{2+}}
\mid \textcolor{blue}{[K,f]^{r\text{-}2+}}
\,.
$$

To conclude, we have seen various formalisms for describing string to string
transformations.  With increasing expressive power, we have sequential functions
(1DFT), rational functions (f1NFT or 1UFT or 1DFT with regular look-ahead or
simple RTE), and regular functions.  Each class of functions is closed under
composition and its equivalence problem is decidable.  The robust and expressive
class of regular functions can be described with various machine models such as
2DFT or 2UFT or f2NFT or SST. It also admits compositional descriptions based on
regular combinators.  We believe that using function composition instead of the
technically involved 2-chained Kleene iteration makes the descriptions much
easier.  Hence, we advocate the use of chr-RTEs or cdr-RTEs as described above.


\end{document}